# Room-temperature Tunable Fano Resonance by Chemical Doping in Few-layer Graphene Synthesized by Chemical Vapor Deposition

Zhihong Liu,† Xiaoxiang Lu,† Peng Peng,† Wei Wu, Steven Pei, Qingkai Yu* and Jiming Bao*

Department of Electrical and Computer Engineering, University of Houston

Houston, TX 77204 USA

ABSTRACT A Fano-like phonon resonance is observed in few-layer (~3) graphene at room temperature using infrared Fourier transform spectroscopy. This Fano resonance is the manifestation of a strong electron-phonon interaction between the discrete in-plane lattice vibrational mode and continuum electronic excitations in graphene. By employing ammonia chemical doping, we have obtained different Fano line shapes ranging from anti-resonance in hole-doped graphene to phonon-dominated in n-type graphene. The Fano resonance shows the strongest interference feature when the Fermi level is located near the Dirac point. The charged phonon exhibits much-enhanced oscillator strength and experiences a continuous red shift in frequency as electron density increases. It is suggested that the phonon couples to different electronic transitions as Fermi level is tuned by chemical doping.

†These authors contribute equally to this work. *To whom correspondence should be addressed.

Graphene has attracted considerable interest for basic and applied research. Not only has graphene provided us with a unique arena to exploit theoretically and experimentally a variety of exotic properties of relativistic mass-less Fermion, but it also enables numerous promising applications such as high-speed transistors, ultrasensitive sensors, super-capacitor, superconductivity, and thermoelectric and optoelectronic devices [1]. Among its many unique properties is the strong electron-phonon interaction, which leads to the breakdown of the adiabatic Born-Oppenheimer approximation and the creation of hybrid electron-phonon mode[2-4]. The effect of this coupling includes the electrical field-induced modification of phonon frequency and line width observed by Raman scattering[2, 5, 6], as well as Fano-like phonon resonance observed with infrared spectroscopy[3, 4]. However, much research is still needed to elucidate the complicated nature of electron-phonon interaction in graphene, especially in the case of Fano resonance, as the results from the two groups are qualitatively quite different from each other, although both groups used high-quality, mechanically exfoliated graphene[3, 4].

Recent advances in chemical synthesis of graphene have paved the way for commercial device applications of graphene[7-11]. Fourier transform infrared spectroscopy (FTIR), a sensitive and versatile optical technique, has been widely used in order to monitor synthesis processes and to verify the final production of graphene. However, the fingerprint of graphene—i.e., the in-plane vibrational mode, has not been clearly identified [10, 12, 13]. This lack of phonon signature in infrared spectroscopy is not necessarily due to the infrared inactive nature of phonon in monolayer graphene. In fact, the same phonon was frequently observed in single-walled carbon nanotubes (SWNTs)[14, 15]. Furthermore, it is very likely that chemically synthesized graphene samples contain few-layer graphene[10, 12, 13], in which the phonon is infrared active[16-18].

Because C-C bonds are the backbone of two-dimensional networks of carbon atoms, their identification is crucial to the characterization and device applications of graphene.

Fig. 1A shows FTIR spectra of few-layer graphene near the infrared active in-plane vibrational mode of graphite (Fig. 1F). The asymmetric line shape is similar to Fano resonance observed in gated double-layer graphene[3, 4], and this line shape is quite different from the absorption band of C=O vibrational mode at ~1730 $cm^{-1}$[12]. The measurement was performed using a multi-bounce attenuated total reflectance (M-ATR) crystal at room temperature (Fig. 1B), where graphene was transferred to the top or both surfaces of the ZnSe crystal in order to increase the interaction length between light and graphene. All of the graphene samples in this work were synthesized by chemical vapor deposition on copper substrates at ambient pressure[19], as opposed to low pressure, which is suitable to grow monolayer graphene[8]. Raman spectra indicate that the sample is mainly made of 3-layer graphene (see supporting documents)[8, 20], although random spots of thicker graphene can be observed. The C=O signal comes mainly from chemical residues of polymethyl methacrylate (PMMA), which is used in transferring graphene from copper substrate to ZnSe ATR[8, 19, 21].

In order to understand the origin of this abnormal phonon line shape, we measure and calculate the phonon spectra of graphite under similar configurations. Fig. 1C shows measured phonon spectra of highly ordered pyrolytic graphite (HOPG) using a single-bounce ZnSe ATR configuration as shown in Fig. 1D[18]. Figure 1E shows the corresponding ATR spectra of HOPG and representative few-layer (5-layer) graphene calculated based on the matrix-transfer method and the optical constants of HOPG[22, 23]. As can be seen, the calculated spectrum of HOPG captures features of both out-of-plane and in-plane infrared active modes[18]. However, the result for few-layer graphene (<100) shows very different spectral characteristics. First, the in-plane $E_{1u}$ phonon at 1589 $cm^{-1}$ of few-layer graphene shows a dip instead of a peak, as for HOPG. Second, the out-of-plane $A_{1u}$ mode at 869 $cm^{-1}$ does not exhibit a well-defined peak. Our calculation further shows that the line shape of in-plane phonon begins to transform from a dip to a peak at about 100 layers, while it is not until more than 180 layers that the out-of-plane phonon spectrum begins to appear as a negative peak. This thickness-dependent phonon line shape can be understood in the following way. For bulk HOPG with the ZnSe ATR technique, there is no internal reflection at the interface between ZnSe and graphite because of the smaller index of ZnSe, and such an obtained spectrum is similar to reflection spectroscopy[18]. However, for few-layer graphene, the total internal reflection happens at the interface between graphene and air, and light passes through graphene twice for each reflection, so the collected light can be treated as a transmitted beam; this is why the few-layer ATR spectrum is similar to the transmission spectrum of thin film, as is shown by the nearby C=O absorption band.

Based on the above results and discussions, we can conclude that the abnormal phonon line shape obtained by the ATR technique is due to the Fano resonance of $E_{1u}$ phonon. On the other hand, we do not observe a clear absorption line shape for the out-of-plane vibrational mode. This result agrees with the calculation, implying that $A_{1u}$ phonon does not have a strong interaction with electron. Fano resonance has been shown to be tuned by gate voltage[3, 4]. Here, we demonstrate the tuning of Fano resonance by chemical doping. Figure 2 shows the change of phonon intensity and Fano line shape when $NH_3$ is used to progressively turn graphene from p-type to n-type[24]. The broadband background has been removed as a third-order polynomial determined outside the phonon peak[3, 4]. The spectrum evolves from anti-resonance to strong interference to phonon dominated as graphene is changed from heavily hole-doped to heavily electron-doped. The observed Fano line shape can be described by the following equation[3, 25]:

$$T(\omega) = T_e \bullet [1 - \frac{(q \bullet r + \omega - \Omega)^2}{(\omega - \Omega)^2 + r^2}] \ ,$$

where Ω is the frequency of coupled or charged phonon, r is the phonon line width, and q is a dimensionless parameter that depends on the coupling between phonon oscillation and continuum transitions as well as their relative strengths. As discussed before[3], q is a negative number, and a Lorentzian absorption line shape is recovered as |q| >> 1. $T_e$ describes the oscillator strength of the charged phonon and is propositional to the infrared effective charge of the phonon [4].

The dependencies of Fano parameters on Fermi level are summarized in Fig. 3A–D. It can be seen that all of these parameters behave quite differently from those reported in previous double-layer graphene[3, 4]. Ω and q experience monotonic change as Fermi level rises from below to above the neutrality point: the magnitude of q increases from much less than 1 to greater than 1; the frequency of coupled phonon decreases from about 2 $cm^{-1}$ above the phonon frequency of graphite to ~3 $cm^{-1}$ below. These behaviors are different from the results of the Manchester group, who observed the highest phonon frequency and the smallest magnitude of q as Fermi level moves close to the neutrality point[4]. Figure 3D shows that the line width of coupled phonon is greater than that of HOPG (~3 $cm^{-1}$), but it fluctuates between 6 and 10 $cm^{-1}$, and there is no significant increase near the neutrality point. In contrast, both groups have shown a much-enhanced line width when phonon is in resonance with continuum transition[3, 4]. Finally, we observed that the oscillator strength $T_e$ becomes slightly weaker as carrier concentration decreases, in agreement with the result of Manchester[4]. However, $T_e$ is much larger in n-doped than p-doped graphene, which is different from that of the Manchester group[4].

The new behavior of Fano resonance and phonon intensity probably stem from new band structure and electron-phonon interaction in 3-layer graphene. As shown in Fig. 4C–E, there are different continuum transitions that have overlap in energy with phonon when the Fermi level is below, close to, or above the neutrality point, so there is always a strong coupling between phonon and electron-hole excitations. On the other hand, there is a clear asymmetry between hole-doped graphene and electron-doped graphene, as is reflected on the behavior of Ω and $T_e$. The continuous red shift of Ω and enhanced strength of $T_e$ in increased electron doping indicate that phonon is coupled to the transitions between bands 4, 5 and band 6 (Fig. 4E)[4, 26]. Similar phonon behaviors have been observed in electron-doped alkali-$C_{60}$ compounds and is attributed to the charged-phonon effect[4, 27, 28]. A number of chemicals other than $NH_3$ and $NO_2$ have recently been reported to make graphene n-type or p-type[29]. $Br_2$ and $I_2$ have been shown to p-dope graphene to a level that cannot be achieved by the electrode gating[30]. It is anticipated that more pronounced phonon intensity will be observed in more heavily doped graphene.

The doping-sensitive Fano resonance also helps to shed light on the doping mechanism of $NH_3$[24, 31]. Theoretical calculation shows that n-doping is due to the charge transfer from adsorbed $NH_3$ to graphene, and that $NH_3$ molecules are oriented on the graphene surface in such a way that H atoms are pointed away from the graphene surface[32]. This picture is supported by our observation. Figure 4A–B shows a series of Fano resonance spectra as $NH_3$ molecules diffuse away from the graphene surface. We can observe a direct correlation between electron density and the absorption band centered at ~3200 $cm^{-1}$. This broad absorption band is due to the N-H stretching modes of $–NH_3$[33].

In conclusion, we have observed optical phonon spectrum and Fano-like phonon resonance in few-layer CVD graphene using room temperature infrared spectroscopy. We have also shown that the Fano line shape and phonon oscillator strength can be controlled by chemical doping, but with different behaviors than previously reported in double-layer graphene[3, 4]. The concurrent FTIR measurement of doping-dependent phonon spectrum and adsorbed chemical species can be used to understand the sensing and doping mechanisms of other chemicals adsorbed on graphene. Our findings have revealed the complicated nature of electron-phonon interaction in few-layer graphene and have paved the way for novel device applications of large-area chemically synthesized graphene.

Acknowledgements

Financial supports from the National Science Foundation under NSF/DMR-0907336 and from the Grants to Enhance and Advance Research (GEAR) of the University of Houston are greatly acknowledged.

Supporting Information Available: This material is available free of charge via the Internet at http://pubs.acs.org.

**Figure captions.**

Figure 1. (Color online) (A) Infrared spectrum of in-plane $E_{1u}$ vibrational mode measured with multi-bounce, attenuated total reflectance (M-ATR) configuration. Graphene was transferred to both the top and bottom surfaces of the ZnSe crystal of an M-ATR in order to increase the absorption signal, and measurement was performed in purging nitrogen. (B) Schematic of ZnSe multi-bounce attenuated total reflection (M-ATR) configuration. Graphene is transferred only to the top surface of ZnSe for the controlled $NH_3$ chemical doping. The $NH_3$ chamber has controlled flow input and output, and it can be detached from the M-ATR. (C) FTIR spectrum of highly ordered pyrolytic graphite (HOPG) with ZnSe single-bounce ATR. (D) Schematic of ZnSe single-bounce attenuated total reflectance (ATR) for HOPG. (E) ZnSe single-bounce ATR phonon spectra of HOPG and 5-layer graphene calculated using optical constants of HOPG. (F) Infrared-active in-plane lattice vibrational mode of double-layer and thicker graphene[17].

Figure 2. (Color online) Evolution of phonon line shape and intensity as graphene is progressively transformed from p-type to n-type by $NH_3$ chemical doping. The doping level is determined by $NH_3$ flow rate and flow duration. The red curves are fits, with Fano parameters and approximate Fermi levels shown in each graph.

Figure 3. (A–D) Fano parameters as a function of Fermi level. Fermi levels are determined by carrier concentrations, which are measured by the Hall effect. Fano parameters are obtained from fits in Fig. 2. Dashed lines in (B) and (D) indicate the position and line width of phonon in HOPG, which are obtained from data in Fig. 1C.

Figure 4. (Color online) (A) FTIR spectral evolution of Fano resonance and $NH_3$ molecules adsorbed on graphene. Graphene is attached to both ZnSe M-ATR surfaces and is initially exposed to pure $NH_3$. FTIR measurement starts after M-ATR is taken out of the $NH_3$ chamber and $NH_3$ is allowed to leave the graphene surface into the air. (B) Close-up of the absorption band of adsorbed $NH_3$ near 3200 $cm^{-1}$. (C–E) Continuum of electronic transitions that could be resonantly coupled to phonon excitation when Fermi levels are at (C) $E_F \sim -E_\Omega/2$, (D) $E_F \sim 0$, and (E) $E_F \sim E_\Omega/2$.

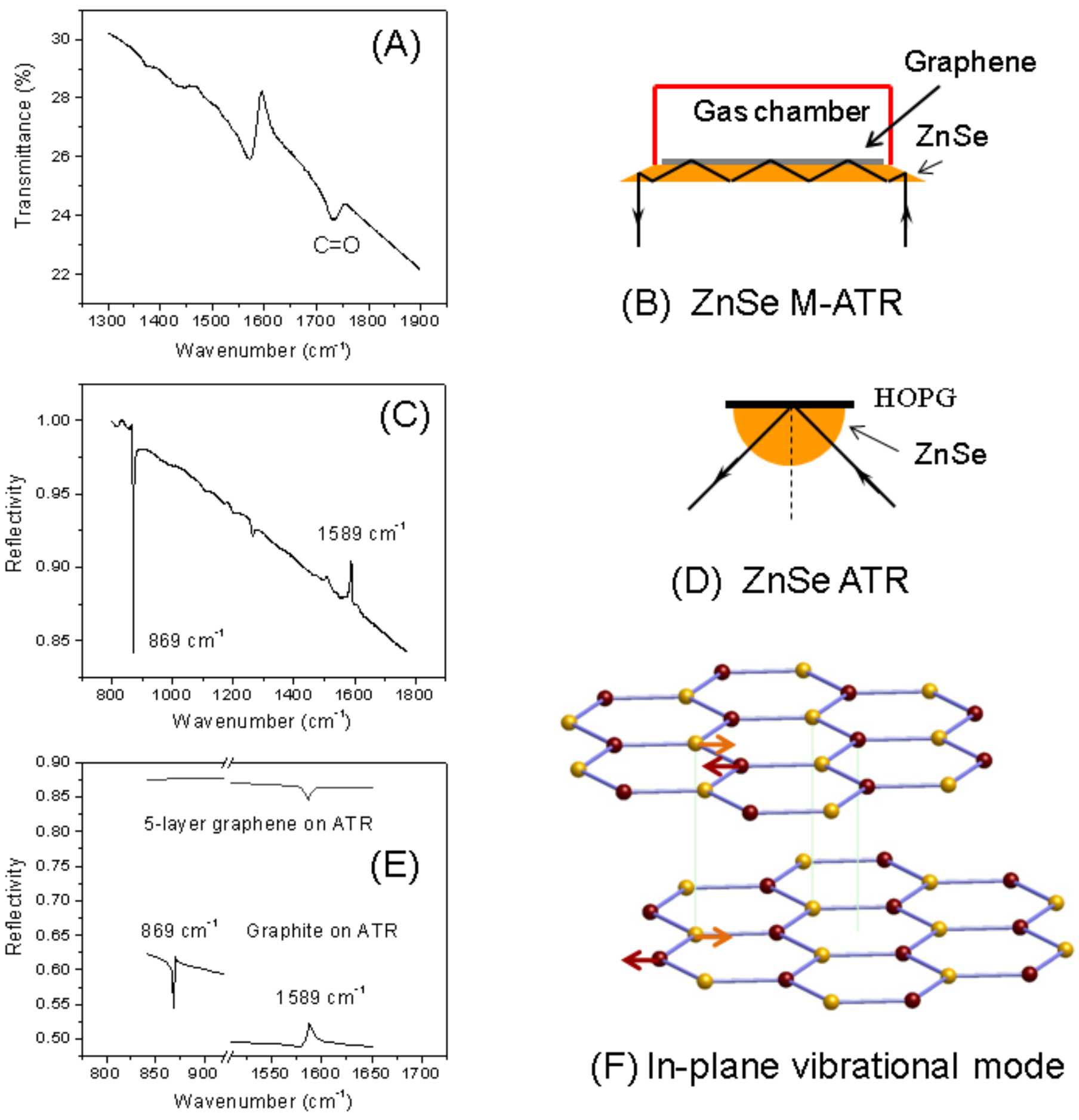

Figure 1

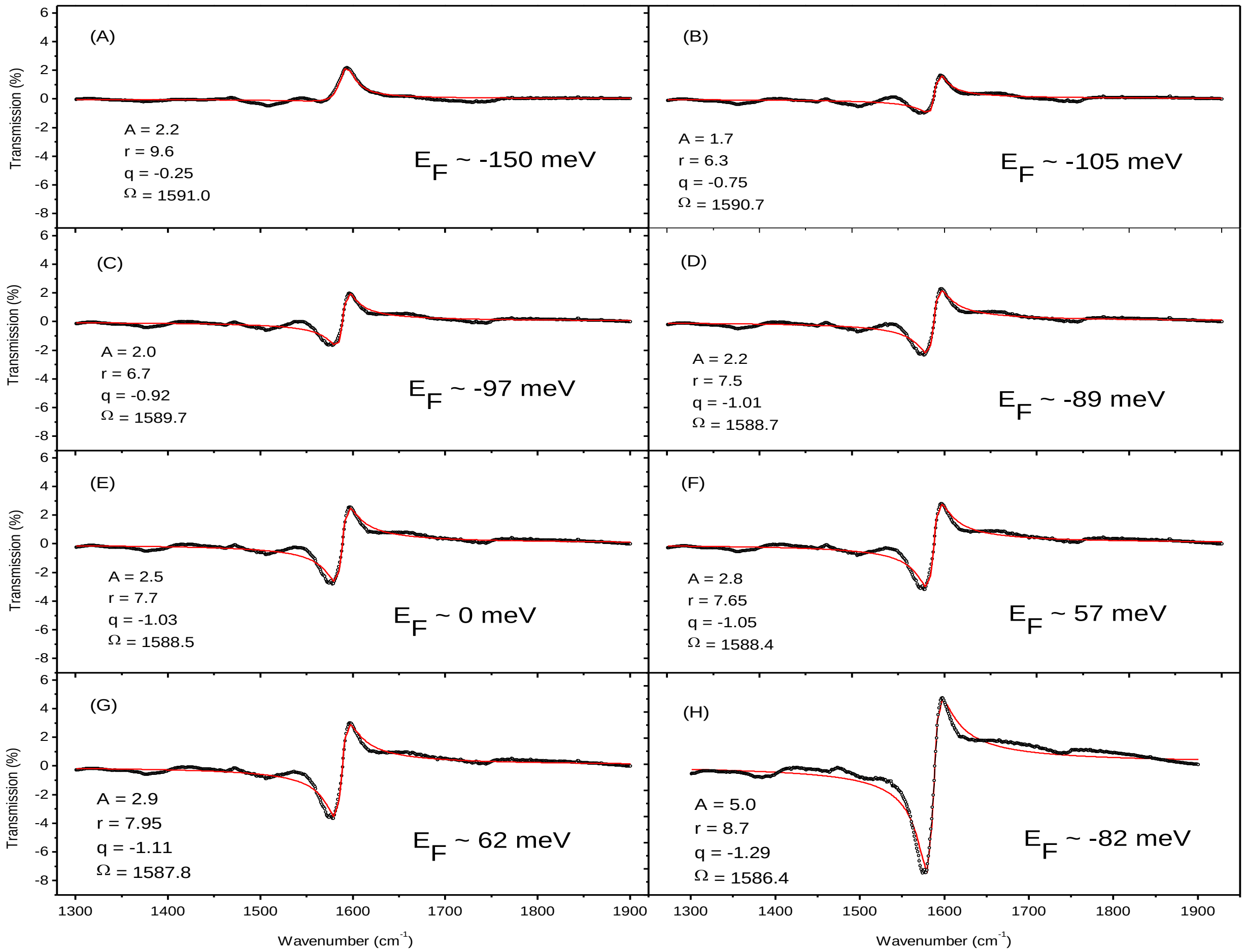

Figure 2

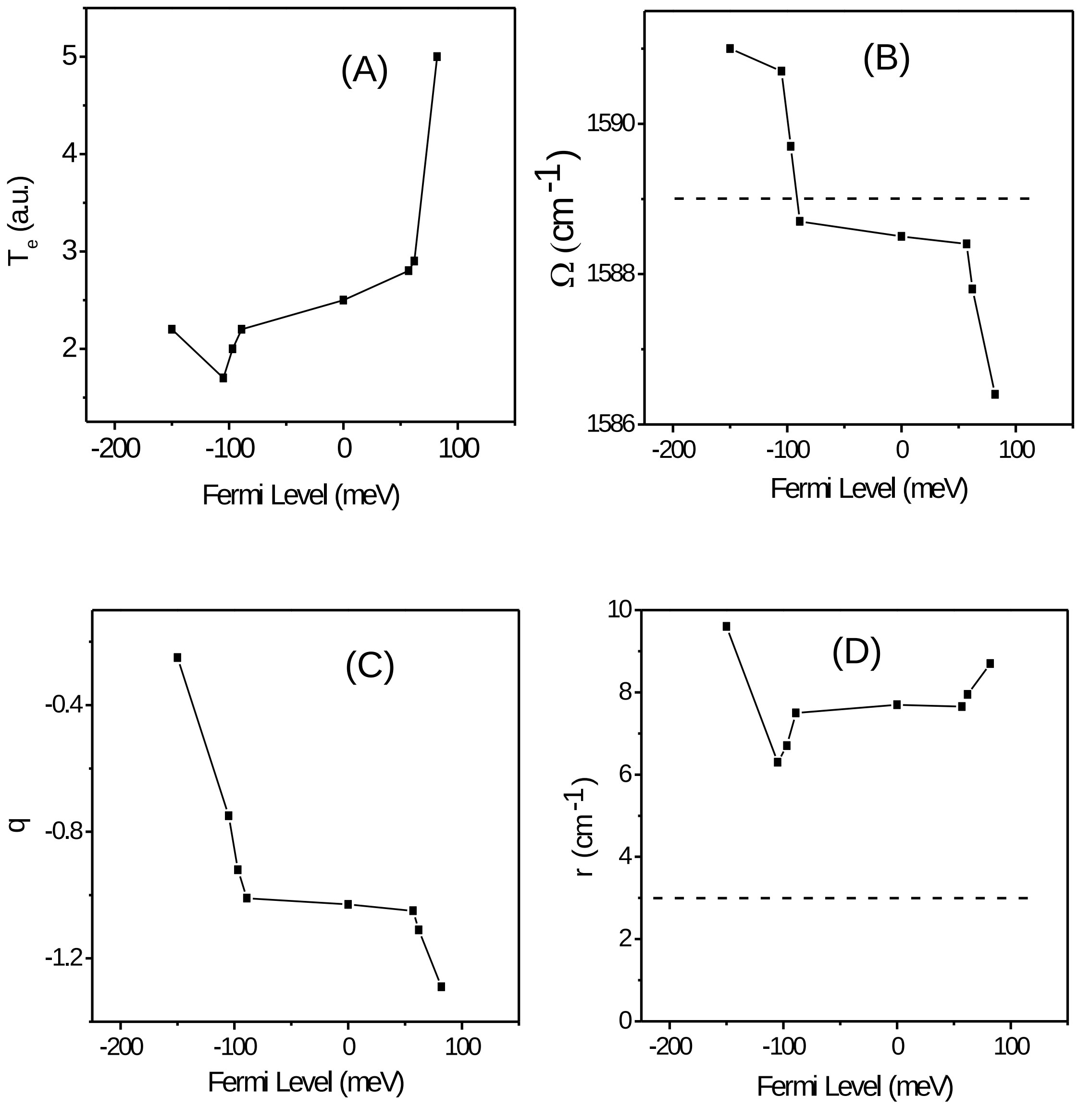

Figure 3

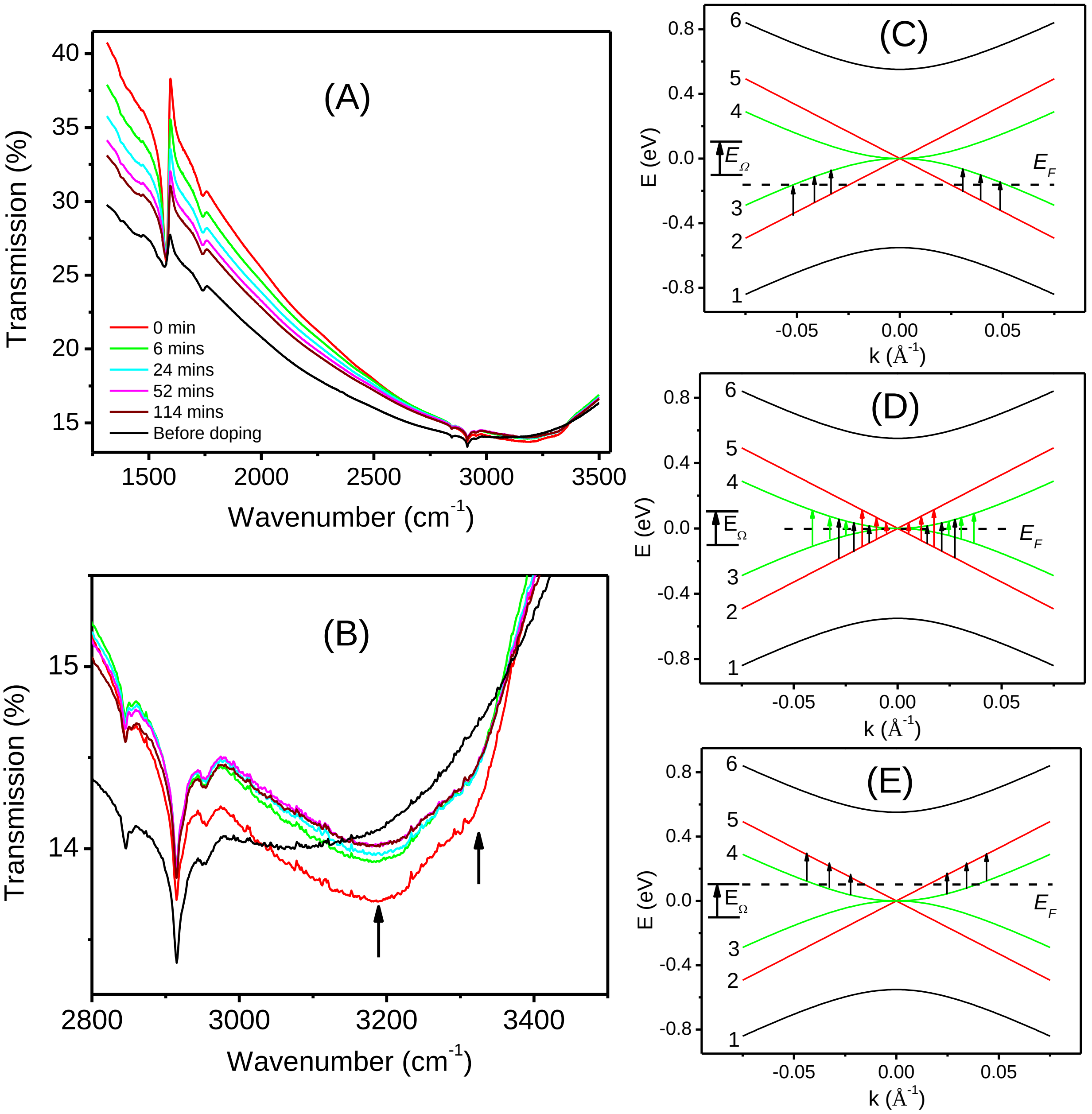

Figure 4